# Luminescence on the horseshoe nanolaser

A. Bogdanov [a], I. Fedorov [b], A.N. Lagarkov [c], G. Tartakovsky [d], A. K. Sarychev [*c]

[a] Hitachi Global Storage Technologies, San Jose Research Center, San Jose, USA,
[b] Moscow Institute of Physics and Technology, Moscow, Russia,
[c] Institute for Theoretical and Applied Electromagnetics RAS, Moscow, Russia,
sarychev_andrey@yahoo.com, [d] Innovative Optical Solutions, San Diego, USA

## ABSTRACT

In this work, we consider a set of ordered split-ring resonators in the form of horseshoes arranged on the layer of active dielectric medium. Size of considered nanoparticles is much less than the wavelength. On this scale, the laser pumping can be treated as a random external force because of the incoherent quantum-mechanical process of photon absorption in the active medium. Then the operation of the plasmonic nanoresonator immersed in the active medium is the response to the action of random force. We find luminescence line and predict its evolution to the laser spectrum when inversion of the active medium is increased to the threshold value. The resonance behavior of the horseshoe can be conveniently tuned by variation of its size and shape. Results are compared with the full quantum mechanical simulation.

**Keywords:** SRR, horseshoe, random force, luminescence, lasing

## 1. INTRODUCTION

The surface plasmon excitations in an ordered array of metal nanoparticles have attracted significant attention in recent years due to the numerous potential applications in the nanoplasmonics[1]. Among the most challenging are electromagnetic metamaterials and devices for transmitting and processing optical signals on a scale much smaller than the wavelength. The unit cell for such plasmonic devices is basically the same – a metal nanoparticle, supporting one or several plasmonic modes, usually surrounded by the dielectric doped with gain medium for loss compensation. In the experiment, usually, a set of such units is excited by external radiation. Then, spectrum of the scattered light is observed. Recent experimental results show remarkable achievements in creating active plasmonic units: authors of the work[2] experimentally obtained generation in the solution of gold nanoparticles with dye-doped shells; paper[3] report on the strongly enhanced luminescence in the pumped nanoantenna-dye system. We are interested exactly is this, sub-threshold regime.

As far as we know, there is no theory describing luminescence in such nanosystems. Usual laser equations (see works[4,5]) result only in zero solution below the lasing threshold, while the "toy model"[6] describes only the passive response of such system to the probe light. Homogeneous laser equation system in Heisenberg approach[4,5] does not contain spontaneous emission noise, which is physically reasonable for the "macroscopic" lasers, when we are only interested in the many photon mode. In the plasmon nanolaser, when photons are replaced by plasmons, the situation is quite the contrary. Number of plasmons cannot be macroscopic due to thermal and strength limitations. Actually, single metal nanoparticle typically has tens of plasmons, and a nanolaser operates in noise-like fashion. «Developed spasing[5]» regime, when plasmon annihilation operator can be treated as c-number, is thus not an appropriate parallel with classical laser theory. Moreover, number of the excited chromophores per nanolaser is also not large (typically, ≲ 100), and pumping process is another important cause of the inhomogeneity in the laser equations. In the Heisenberg picture, these disturbances can be taken into account by supplementing system of equations with Langevin noise operators[7]. Though, problem of solving system of operator equations will remain.

In the present work, we propose another approach that allows obtaining spectrum of luminescence of the system plasmonic nanoparticle-laser medium. In our model, the excited chromophore (dye molecule, quantum dot, etc.) transmits its energy to the plasmonic modes via the nearfield interaction. The chromophore is excited, in turn, in the random moments by optical or electric pumping. Thus, in our model nanoresonator responds to the random external force. We obtain non-trivial solution for any pumping intensity and trace the transition to the lasing.

## 2. METHODS

**2.1 Plasmon equation**

One of the promising plasmonic nanodevices is split-ring resonator (SRR) due to its relatively simple fabrication by electron-beam lithography. SRR are usually placed in the ordered array on the active substrate. Real experiment structures usually look like "plane horseshoe" on the Fig. 1, but we will consider its "volume" counterpart (Fig. 1), is mathematically simpler. We shall return to the plane horseshoe geometry later.

Volume horseshoe is similar to the plane capacitor, with gain medium placed between its plates. Dimensions of the volume horseshoe in Z and Y directions are much larger than distance between plates. Thickness of these plates is small compared to the plasmon wavelength and distance between plates. Current density in the plate is constant in the X direction.

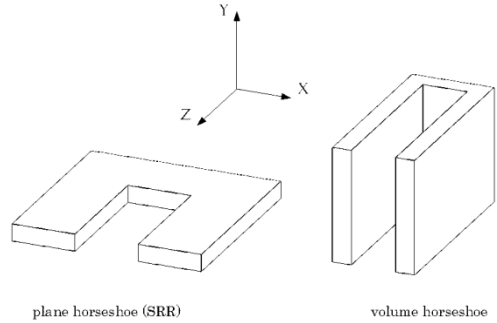

Figure 1. Model geometry.

There are several plasmonic modes detected in experiments on the excitation of plasmons in SRR (with geometry close to our model) by the resonance light[8]. These modes can be divided into the two types: symmetrical and antisymmetrical, when electrical current in the horseshoe arms is directed in the same and opposite direction, correspondingly. First type does not produce any significant magnetic dipole moment in the SRR, whereas the second type does. This feature is crucial for creating artificial magnetic moment (and, thus, metamaterials with negative magnetic permeability), and we focus in the present work only on the second type modes.

FOM of the horseshoe resonator in near infrared and even optical range is typically $\sim 10 - 100 \gg 1$, and plasmon eigenmodes can be found with the same technic as in the "classical" waveguide. Physically, these modes are similar to standing waves in the waveguide line formed by two parallel metal plates (Z direction on fig. 1), closed by the bridge at one endings (fig. 2). Bridge is a short segment of the same metal strip. It inevitably disturbs the electric current in the strip line; in our model we neglect this effect and suppose that current in the strips is parallel to Z axis. In our model, bridge can be taken into account by appropriate boundary conditions for the current.

If the wavelength in much more than transversal size of the waveguide, mode has form of the principal wave [9], when electric field is transversal to the wave direction $Z$. For the first antisymmetric mode we have $\lambda/4 \approx a \approx 3d$, and condition is met well; at some higher harmonic it becames violated, and calculation of EM fields becames much more complicated. In such principal wave, spatial EM field distribution is the same as in case of stationary charges and currents. That is, we consider plane capacitor, where current, electric and magnetic fields are directed along Z, X and Y axis correspondingly.

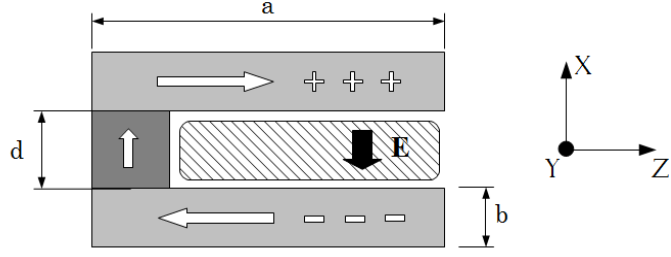

Figure 2. Scheme of the volume horseshoe. Bridge area is denoted by dark gray color. Cross-hatched area denotes active medium. White rows indicate current directions in the antisymmetric mode.

Following[10], we use the Ohm's law for derivation of the equation for the current in the SRR:

$$\boldsymbol{J}Z = \boldsymbol{E} = -\nabla\varphi + ik\boldsymbol{A} \quad (1.1)$$

, where $\boldsymbol{A}$ and $\varphi$ are vector and scalar potentials of the electric field $\boldsymbol{E}$, $k = \omega/c$, $\boldsymbol{J}$ is current density in the plate and $Z$ is surface impedance of the plate. If plates are much thinner than the skin depth, we have $Z = (\sigma b)^{-1}$, where $\sigma$ is conductance of metal.

As we are interested only in the antisymmetric current modes, charge, current and potentials take opposite values at points with opposite $x$ on the different horseshoe's arms. Then, equations 1.1 for different plates are equivalent and we proceed with equation only for the upper plate. In these assumptions, Eq. (1.1) transforms to

$$JZ = \left(-\frac{\partial}{\partial z}\varphi + ikA\right)\bigg|_{x=d/2} \quad (1.2)$$

We have two sources of the potentials: charges at the plates and charges on the chromophores. Charges at the plates contribute as in the plane capacitor, and we can write

$$\varphi|_{x=d/2} = \frac{Q}{2C_0} + \varphi^{chr}|_{x=d/2}, \quad C_0 = \frac{1}{4\pi d},$$
$$A|_{x=d/2} = \frac{L_0 J}{2} + A^{chr}|_{x=d/2}, \quad L_0 = \frac{4\pi d}{c}, \quad (1.3)$$

where $Q$ and $C_0$ are electric charge and capacity of the plane capacitor per unit area of the plate. $L_0$ is specific inductance of the two parallel plates. When volume between plates is filled with the host dielectric, it can be easily shown that charge induced on the surface of the dielectric equals

$$Q_{ind} = -\frac{\epsilon_d - 1}{\epsilon_d} Q, \quad (1.4)$$

where $\epsilon_d$ is dielectric permittivity. This induced charge sums up with charges on the metal and effectively changes capacity and inductance: $C_0 \to \epsilon_d C_0$, $L_0 \to L_0/\epsilon_d$. Dielectric screens charges on the chromophores in the same fashion, so that $\varphi^{chr} \to \varphi^{chr}/\epsilon_d$, $A^{chr} \to A^{chr}/\epsilon_d$. In effect, in the presence of the host dielectric, right side of Eq. (1.2) with the same potentials 1.3 should be multiplied by $\epsilon_d$. Then, substituting 1.3 into 1.2 and using charge conservation law in the form

$$i\omega Q = \frac{\partial J}{\partial z}, \quad (1.5)$$

we obtain

$$\frac{1}{i\omega C_0}\frac{\partial^2 J}{\partial z^2} + (2\epsilon_d Z - ikL_0)J = 2\left(-\frac{\partial}{\partial z}\varphi^{chr} + ikA^{chr}\right)\bigg|_{x=d/2} \quad (1.6)$$

Without the right side it is equation for current in the free horseshoe. Together with boundary conditions, it defines coordinate eigenfunctions for current. In the simplest case, when impedance of the bridge between arms is zero, boundary conditions are

$$J|_{z=a} = 0, \qquad \frac{\partial J}{\partial z}\bigg|_{z=0} = 0 \tag{1.7}$$

In more general case of arbitrary bridge impedance $Z_b$, second boundary condition is

$$\frac{Q|_{z=0}}{C_0} = \frac{1}{i\omega C_0}\frac{\partial J}{\partial z}\bigg|_{z=0} = dZ_b J|_{z=0},$$

and eigenfunctions became more awkward. For simplicity, we will proceed with conditions 1.7. We consider modes when $Q$ and $J$ are functions of one coordinate $z$ (independent of $y$). Then, 1.7 and 1.5 give sets of orthonormal functions for current and charge on the plate area $A = \{0 < y < l\} \oplus \{0 < z < a\}$:

$$R_{Jn}(y,z) = \left(\frac{2}{al}\right)^{1/2} \cos(k_n z), \qquad J = \sum_n J_n(\omega) R_{Jn}(y,z)$$

$$R_{Qn}(y,z) = \left(\frac{2}{al}\right)^{1/2} \sin(k_n z), \qquad Q = \sum_n Q_n(\omega) R_{Qn}(y,z) \tag{1.8}$$

$$k_n = \frac{\pi}{a}\left(n + \frac{1}{2}\right),$$

which we will use for separation of variables in 1.6. Note, that Eq. (1.6) takes into account both electic and magnetic impact of the active medium on the plasmonic mode. Electric part, proportional to $\varphi^{chr}$, can be expressed as

$$\varphi^{chr}(r) = \int_{V_1} \frac{\rho(r_1)}{|r-r_1|} dr_1 = \int_{V_1} \frac{-div\boldsymbol{P}(r_1)}{|r-r_1|} dr_1, \tag{1.9}$$

where $\rho$ is density of microscopic charges on the dye molecules, $V_1$ is the volume containing the horseshoe, $\boldsymbol{P}$ is macroscopic polarization of the ensemble of the chromophores, defined as

$$\boldsymbol{P}(\boldsymbol{r}) = \sum_i \delta(\boldsymbol{r} - \boldsymbol{r}_i)\boldsymbol{d}_i, \tag{1.10}$$

where $\boldsymbol{d}_i$ is dipole moment of the i-th chromophore. (here, like everywhere in this paper, we neglect delay effects because all distances in the system are much less than the wavelength in the dielectric). We expand $\varphi^{chr}$ on the upper plate as

$$\varphi^{chr}|_{x=d/2} = \sum_n \varphi_n^{chr}(\omega) R_{Qn}(y,z) \tag{1.11}$$

Then,

$$\varphi_n^{chr} = \int_A R_{Qn} dr_{yz} \int_{V_1} \frac{-div\boldsymbol{P}(r_1)}{|r-r_1|} dr_1 \bigg|_{r_x=d/2} = -\int_{V_1} div\boldsymbol{P}(r_1) dr_1 \int_A \frac{R_{Qn}(y,z)}{|r-r_1|} dr_{yz}\bigg|_{r_x=d/2} \tag{1.12}$$

Surface integral in the second line 1.12 is the value of the scalar potential, produced by unit charge distribution $R_{Qn}$ on the upper plate of the n-th plasmon mode. In the antisymmetric mode this value is half of the full potential value produced by both plates. Denoting the latter as $\varphi_{n1}$ and integrating by parts, we find

$$\varphi_n^{chr} = -\frac{1}{2}\int_{V_1} div(\boldsymbol{P})\varphi_{n1} d\boldsymbol{r} =$$

$$= -\frac{1}{2}\oint_{\partial V_1} (\boldsymbol{P}\varphi_{n1})\boldsymbol{n} df + \frac{1}{2}\int_{V_1} \mathbf{grad}(\varphi_{n1})\boldsymbol{P} d\boldsymbol{r} = -\frac{1}{2}\int_{V_1} \boldsymbol{P}\boldsymbol{E_{n1}} d\boldsymbol{r} \tag{1.13}$$

, where $\partial V_1$ is boundary of the volume $V_1$, which we can take large enough so that surface integral vanishes. $\boldsymbol{n}$ is the outer normal of the surface. In the last line, we introduced $\boldsymbol{E_{n1}} = -\mathbf{grad}\,\varphi_{n1}$.

Magnetic part in the right side of 1.6, proportional to $A^{chr}$, expands exactly in the same way, but functions $R_{Jn}$ should be used instead of $R_{Qn}$. From now, we drop this term for simplicity. Equation 1.6 then reduces to

$$\frac{\partial^2 J}{\partial z^2} + (k^2 + 2i\omega\epsilon_d C_0 Z)J = 4\pi i\omega C_0 \sum_n k_n R_{Jn} \int_{V_1} P_x(\omega, \mathbf{r}) R_{Qn}(\mathbf{r}) d\mathbf{r}, \qquad (1.14)$$

For specific harmonic, Eq. (1.14) gives

$$\left(k^2 - k_n^2 - \frac{2\epsilon_d}{\epsilon_m b d}\right) J_n(\omega) = \frac{i\omega}{d} k_n P_n(\omega), \qquad (1.15)$$

where we used $Z = \frac{1}{\sigma b} = \frac{4\pi i}{\epsilon_m \omega b}$, $C_0$ from Eq. (1.3), and denoted

$$P_n(\omega) = \int_{V_1} P_x(\omega, \mathbf{r}) R_{Qn}(\mathbf{r}) d\mathbf{r} \qquad (1.16)$$

is gain polarization, projected onto the n-th charge eigenmode. Eq. (1.15) without the right hand side determines the resonant frequencies of different plasmonic modes.

Note, that Eq. (1.14) is valid not only for geometry as in Fig. 2. Actually, it holds for any waveguide line with multiply connected cross-section, with appropriate capacitance $C_0$ and impedance $Z$. Constant $C_0$, calculated in the plane capacitor approximation, is actually valid when plasmon wavelength $\lambda \gg d$. For the real geometry and main plasmonic mode ($n = 0$) we have $\lambda/4 = a \approx 3d$ and condition $\lambda \gg d$ holds well. For higher harmonics, $C_0$ depend on the spatial charge distribution, and, therefore, on the mode index.

**2.2 Polarization equation**

Individual chromophores contribute in Eq. (1.14) via polarization $\mathbf{P}$ according to 1.10. In our model, we represent them as effective two-level systems (TLS), with dipole moments proportional to the non-diagonal elements of the density matrix $\boldsymbol{\rho}$ for i-th chromophore:

$$\mathbf{d}_i = \mathbf{\Pi}_i \rho_{ab}^i + c.c. \qquad (2.1)$$

, where $\mathbf{\Pi}$ is dipole moment corresponding to the laser transition in the chromophore. In the rotating wave approximation, these elements evolve in the electric field $\mathbf{E}$ according to well-known expression[7]

$$\frac{d}{dt}\rho_{ab}^i(t) = -(i\omega_r + \gamma)\rho_{ab}^i - \frac{i}{\hbar}\mathbf{E}(\mathbf{r}_i)\mathbf{\Pi}_i^*(\rho_{aa}^i - \rho_{bb}^i), \qquad (2.2)$$

where $\omega_r$ is the frequency of the transition, and $\gamma$ is its total decay rate.

In usual laser model, pumping of the gain medium leads to appearance of the TLS in the upper state $a$, so that $\rho_{aa} = 1$, $\rho_{bb} = 0$ in the moment of excitation. In sub-threshold regime, electric field in 2.2 is zero and TLS cannot produce any dipole moment. In fact, there always are field fluctuations and chromophores spontaneously emit photons in the laser mode. As we noted earlier, this noise cannot be neglected in the nanolaser. Indeed, in this case spontaneous emission is many orders of magnitude stronger than in usual optical laser due to the trifling modal volume of the nanoresonator; actually, in the weak-pumping luminescence regime it can be the principal source of plasmons in nanoparticle.

According to the concept of quantum jumps[11], transitions between eigenstates of atom occur abruptly at the well-defined moments in time. Such transitions between the metastable energy levels were observed in the experiments on the resonant florescence on the single atom[12]. Transitions between energy levels in the chromophores happen exactly in the same way. Thus, in the pumping and relaxation process, chromophore changes its state (and, consequently, the dipole moment) in discrete steps. Taking it all in consideration, we suggest the following phenomenological equation for the quantity $\rho_{ab}$:

$$\frac{d}{dt}\rho_{ab}^i(t) = -(i\omega_r + \gamma)\rho_{ab}^i - \frac{iD_0}{\hbar}\mathbf{E}(\mathbf{r}_i)\mathbf{\Pi}_i^* + f_i(t), \qquad (2.3)$$

Quantity $f_i$ is an external random force, which has the form

$$f_i(t) = \sum_\mu f_\mu^i \delta(t - t_\mu^i) \tag{2.4}$$

In agreement with the idea of quantum jumps, it represents action of the pumping and decay mechanisms on the i-th chromophore. We think of elementary acts of excitation and relaxation as of instantaneous events, expressed in the form of delta pulses with some complex amplitudes $f_\mu^i$, occuring at moments $t_\mu^i$. Average value of these amplitudes $\langle f_\mu^i \rangle$ is zero.

Concept of the random force naturally implies initiation of the TLS with non-zero dipole moment, or $\rho_{ab}$. It is more realistic then initiation in the pure upper state for several pumping technics[13]. Initial value of $\rho_{ab}^i$ enters the right side of Eq. (1.14) with factors $(R_{Qn}(z_i)\Pi_x^i)$, which is coupling coefficient between i-th chromophore and n-th plasmonic mode. That is, energy of the chromophore can be transmitted even to the "empty" resonator, which represents the spontaneous emission process. Finally, we set $\rho_{aa}^i - \rho_{bb}^i \equiv D_0 = const$. It makes TLS dynamics simpler and allows us to keep equations linear. This assumption neglects depletion of the gain medium which is reasonable since we are interested only in the luminescence regime when electric field is weak. Constant $D_0$ corresponds to the average population inversion of the laser level.

Now, we can derive equation of motion for $P_n$. As long as in the horseshoe electric field $E = 4\pi Q$, using 1.5, 1.10, 2.1 and 2.3, in frequency domain we get

$$(i\Delta - \gamma)P_n(\omega) + D_0 \Omega_n J_n(\omega) + F_n(\omega) = 0 \tag{2.5}$$

, where $\Delta = \omega - \omega_r$ and we have introduced aggregated quantities

$$\Omega_n = \frac{4\pi k_n}{\hbar \omega} \sum_i \left(R_{Qn}(z_i)\Pi_x^i\right)^2 \tag{2.6}$$

$$F_n(\omega) = \int_0^T dt\, F_n(t) e^{i\omega t}, \quad F_n(t) = \sum_i \left(R_{Qn}(z_i)\Pi_x^i\right) f_i(t) = \sum_\nu F_\nu^n \delta(t - t_\nu) \tag{2.7}$$

$\Omega_n$ is dimensionless coupling constant between n-th mode and gain medium. $F_n$ is effective force acting on the n-th mode. $T$ is time passed from the start of exposure of the horseshoe to the pumping radiation.

### 2.3 Solution

Equations 1.15 and 2.5 form the full set of equations for n-th mode in the system. Solutions for $J_n$ and $P_n$ have form

$$J_n = \frac{-F_n(\omega)}{(i\Delta-\gamma)g_n + D_0\Omega_n}, \tag{3.1}$$

$$P_n = \frac{-g_n F_n(\omega)}{(i\Delta-\gamma)g_n + D_0\Omega_n}, \tag{3.2}$$

where we have denoted

$$g_n = \frac{d}{i\omega k_n}\left(k^2 - k_n^2 - \frac{2\epsilon_d}{\epsilon_m bd}\right) \tag{3.3}$$

Experimentally observable intensities are proportional to the quantities $\frac{1}{T}\langle |J_n|^2 \rangle$ and $\frac{1}{T}\langle |P_n|^2 \rangle$, where angle brackets mean averaging over all possible realizations of random force $F_n$. As long as

$$\langle |F_n(\omega)|^2 \rangle = \langle \int_0^T dt_1\, F_n(t_1)e^{i\omega t_1} \int_0^T dt_2\, F_n^*(t_2)e^{-i\omega t_2} \rangle =$$
$$= \int_0^T \int_0^T dt_1\, dt_2 \langle F_n(t_1)F_n^*(t_2)\rangle e^{i\omega(t_1-t_2)} \tag{3.4}$$

, and $F_n(t)$ is delta-correlated stationary process[14], we have

$$\langle |F_n(\omega)|^2 \rangle = G_n \int_0^T \int_0^T dt_1\, dt_2\, \delta(t_1 - t_2)\, e^{i\omega(t_1 - t_2)} = G_n T \qquad (3.4)$$

, where we used $\langle F_n(t_1) F_n^*(t_2) \rangle = G_n \delta(t_1 - t_2)$. Then, we obtain expressions for intensities of light radiated by the oscillating horseshoe current and polarization:

$$I_{Jn} \propto \frac{1}{T} \langle |J_n|^2 \rangle = \frac{G_n}{|(i\Delta - \gamma)g_n + D_0 \Omega_n|^2} \qquad (3.5)$$

$$I_{Pn} \propto \frac{1}{T} \langle |P_n|^2 \rangle = \frac{|g_n|^2 G_n}{|(i\Delta - \gamma)g_n + D_0 \Omega_n|^2} \qquad (3.6)$$

Correlator $G_n$ of the random force 2.7 equals

$$G_n = \langle |F_\nu^n|^2 \rangle R = \langle |R_{Qn}(z_i)\Pi_x^i f_\mu^i|^2 \rangle R \qquad (3.7)$$

, where angle brackets mean averaging over index of chromophores $i$ and realizations of individual random forces $f_\mu^i$. $R$ is average rate of pulses in process $F_n$, or an average number of photons observed by gain medium per time unit.

Correlator 2.7 and coupling constant 2.6 can be estimated as

$$\Omega_n \approx \frac{4\pi N \Pi^2 k_n}{\hbar \omega a l}, \qquad G_n \approx \frac{\Pi^2 R}{3 a l} \qquad (3.8)$$

, where we took $\langle f_\mu^{i\,2} \rangle \approx 1$, and $N$ is number of chromophores per one horseshoe, and $\Pi = \langle |\Pi_x^i| \rangle$. Note, that for the chosen coordinate functions 1.8, estimated correlator 3.8 does not depend on the mode index.

## 3. RESULTS AND DISCUSSION

The time derivative of a polarization $P$ is, in effect, a current of the polarization charges. Then, we may think of the luminescence field as being radiated by this oscillating current. Spectrum of its power is proportional to the quantity 3.6. Expression 3.5 gives the spectrum of the horseshoe radiation, which is produced by the oscillating plasmonic mode. Then, the full spectrum of the scattered light is a linear combination of expressions 3.5 and 3.6. In our case, when the nanoresonator is much smaller than the wavelength, this horseshoe radiation is strongly suppressed by the ohmic loss in metal.

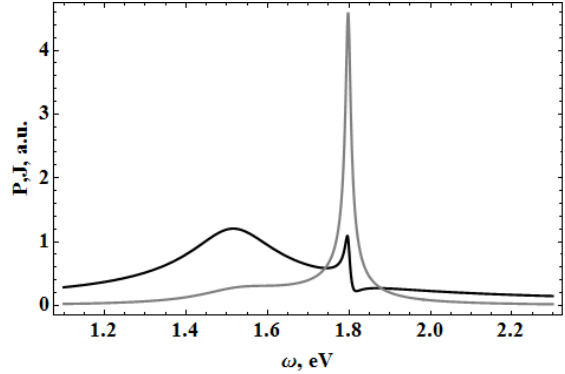

Figure 3. Current 3.5 (gray line) and polarization 3.6 (black line) in arbitrary units at inversion $D_0 = 0.015$. We took number of chromophores $N = 10^3$ and chromophore dipole moment $\Pi = 10^{-16}\ CGS$. Horseshoe parameters are $a = l = 60\ nm$, $b = 10\ nm$, $d = 20\ nm$. Permeability of the host dielectric $\epsilon_d = 10$, $\omega_r = 1.5\ eV$, $\gamma = 0.1\ eV$.

Fig. 3 shows luminescence line at fixed $D_0$ together with spectrum of the horseshoe radiation. For graphics, we took $n = 1$, which corresponds to the first plasmonic mode in the horseshoe. For the metal permittivity, we use Drude model

$\epsilon_m = \epsilon_b - \left(\frac{\omega_p}{\omega}\right)^2 / \left(1 + i\frac{\omega_t}{\omega}\right)$. For silver we took $\epsilon_b = 5$, $\omega_p = 9eV$, $\omega_t = 0.02eV$. Fig. 4 shows polarization spectrum 3.6, plotted versus frequency for different $D_0$:

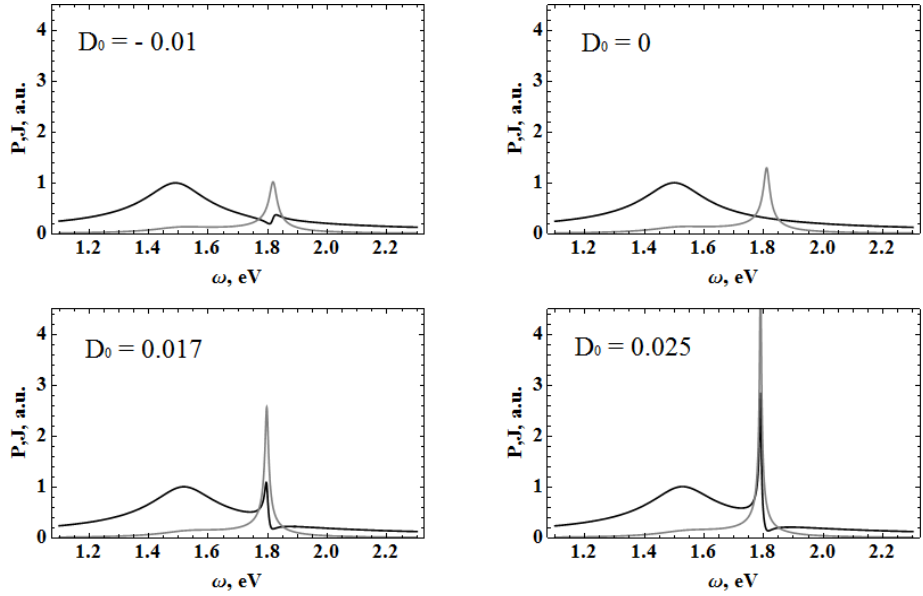

Figure 4. Shape of the luminescence line. Line is formed by the broad Lorentzian line disturbed by narrow horseshoe resonance line. Note that disturbance is zero when $D_0 = 0$.

In absense of the SRR, experimental spectrum will be a pure Lorentz line of the chromophores working transition (fig. 4, black line on the picture with $D_0 = 0$). When SRR is included, chromophores couple to plasmonic mode of the horseshoe and their luminescence spectrum acquires disturbance at the resonance frequency of the nanoresonator. Resonance line of the silver or gold SRR is usually much thinner than transition line of the chromophores. Physically, transition in the active medium and plasmonic mode of the nanoresonator interact as coupled oscillators, and disturbance in the luminescence spectrum is similar to the Fano-resonance picture (fig. 3, 4).

Under the strong pumping when average inversion of the chromophores reaches some critical value, the nanoresonator with the neighboring chromophores form a strongly coupled system called "nanolaser" or "spaser". At some positive inversion solution of eqs. 1.15 and 2.5 diverges and one sees functions 3.5 and 3.6 going to infinity. This point is known as laser threshold. At threshold and beyond electric field cannot be treated as weak, and saturation effects became important. Figure 5 shows this evolution of the luminescence line with growth of inversion up to the onset of lasing.

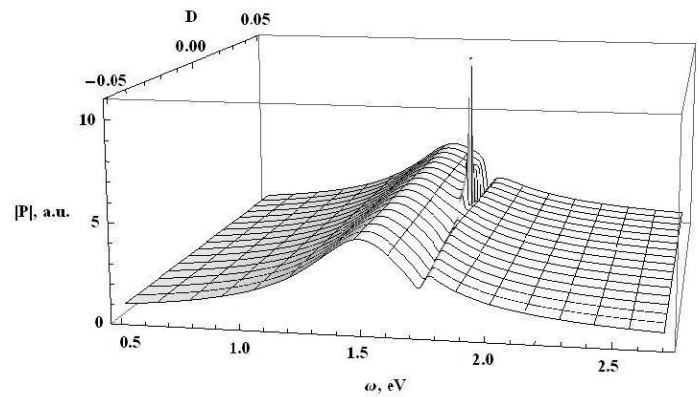

Figure 5. Luminescence polarization spectrum 3.6, plotted versus population inversion $D_0$ and frequency.

Now let us calculate the laser linewidth as a function of the inversion $D_0$ in the simplest case, when the horseshoe is tuned exactly in resonance with the chromophore transition. Resonance frequency of the SRR can be obtained from 1.15. Using Drude model for metal permittivity one gets

$$\omega_n^{horseshoe} \cong ck_n \left(1 + \frac{2\epsilon_d c^2}{bd\omega_p^2}\right)^{-1/2} \tag{4.1}$$

The set of parameters used for figure 3 implies that unity can be dropped in 4.1. Hence, resonance condition takes the form

$$\frac{\omega_r}{\omega_p} \cong k_n \sqrt{\frac{bd}{2\epsilon_d}} \tag{4.2}$$

If condition 4.2 is met, after some algebra one gets from 3.1 or 3.2 an expression for HWHM of the luminescence line:

$$Lw \cong \frac{\omega_\tau}{2} \frac{D_0^{thr} - D_0}{D_0^{thr}}, \quad D_0^{thr} = \frac{V\gamma\omega_t\hbar}{4\pi N\omega_r \Pi^2} > D_0, \tag{4.3}$$

where $D_0^{thr}$ is seen to be a threshold inversion, in compliance with[4]. $V = adl$ is mode volume which is in our case space between the arms of the horseshoe. In agreement with fig. 5, linewidth 4.3 goes exactly to zero at the threshold.

Of course, this idealistic prediction never realizes in measurements. Real linewidth is caused by quantum fluctuations. These are known to be strongest at threshold, when energy of the mode is around that of a single photon[15]. It is extremely interesting to compare semiclassical model presented here with the full quantum mechanical simulation. It will be presented in detail elsewhere, and here we shall confine ourselves only with a brief glimpse into the results for linewidth and number of photons.

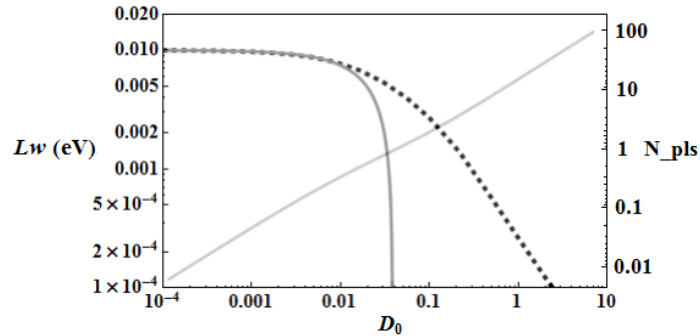

Figure 6. Luminescence linewidth 4.3 (solid dark gray line) and the results of the quantum computation in double logarithmic scale. Thick dotted line is "quantum" linewidth, thin solid line is the number of plasmons (right $y$ axis). Parameters used: $n = 0$, $N = 10^2$, $\Pi = 5 \times 10^{-17}$ CGS, $a = 60\ nm$, $l = 30\ nm\ b = d = 10\ nm$, $\epsilon_d = 1$, $\omega_r = 1.5\ eV$, $\gamma = 0.2\ eV$. Metal parameters are the same. For convenience, pumping rate from the quantum model is reduced to the equilibrium inversion $D_0$. At threshold, quantum theory predicts linewidth and number of plasmons to be $4.6 \times 10^{-3}\ eV$ and $1.15$, respectively.

Figure 6 presents comparison of the "classical" expression for the linewidth 4.3 with prediction of our fully quantum mechanical model for the same parameters of the system. As expected, "quantum" linewidth is not going to zero at the threshold, but falls to only about a half of the initial, "cold cavity" value.

Plasmonic nanoresonator, as we noted earlier, is not expected to support more than tens of plasmons in CW regime. Hence, as appears from the picture, narrowing of the nanolaser linewidth up to $10 - 30$ times seems realistic for CW, and even more for pulsed regime.

The "classic" threshold line 4.3 intersects the number of plasmons line in its nonlinear region, which indicates agreement between our theory and quantum computation. Note that intersection occurs at approximately unity number of quanta, which fits well to the definition of threshold for "thresholdless" microlasers[15].

Fig. 6 shows that our semiclassical model is adequate in its definitional domain except a small region near the threshold. In fact, quantum fluctuations which are strong in that region, are totally negligible elsewhere below threshold. Effectively, usability condition for our model is that number of quanta should be no more than unity.

## CONCLUSION

We propose an original solution of the problem of luminescence in system metal horseshoe adjacent to the gain medium under the external optical pumping in the sub-threshold regime. We predict the form and linewidth of the luminescence line versus the inversion of the gain medium. Comparison of the results with the quantum mechanical computation confirms usability our semiclassical model. Our approach to the classical description of the plasmonic modes in the horseshoe is easily applicable to any waveguide line with multiply connected cross-section. Proposed phenomenological equation for the gain medium dynamics can be applied to broad range of problems in nanoplasmonics. We would like to acknowledge the support from RFBR grant №12-02-01365.